\documentclass[12pt]{article}

\usepackage{amsmath,amsthm,amssymb,ascmac}
\usepackage{bm}
\usepackage[english]{babel}
\usepackage{color}
\usepackage{url}
\usepackage{cite}

\newcommand{\bq}{\begin{equation}}
\newcommand{\eq}{\end{equation}}
\newcommand{\ba}{\begin{eqnarray}}
\newcommand{\ea}{\end{eqnarray}}
\newcommand{\ezero}{\setcounter{equation}{0}}

\newcommand{\p}{\partial}

\newcommand{\ep}{\epsilon}

\newcommand{\aeq}{&\!\!\!=\!\!\!&}

\newcommand{\dl}{\delta}
\newcommand{\lm}{\lambda}

\newcommand{\lab}{\label}
\newcommand{\no}{\nonumber}
\newcommand{\re}[1]{(\ref{#1})}

\newcommand{\bea}{\begin{eqnarray}}
\newcommand{\eea}{\end{eqnarray}}

\newcommand{\RM}[1]{{\rm{#1}}}

\title{Deriving Feynman's Consistency Condition from His Abandoned Energy--Momentum Approach}

\author{
Satoshi Nakajima\thanks{Independent researcher, Japan. subarusatosi@gmail.com} \ and 
Antonio L\'opez-Pinto\thanks{Independent researcher, Spain. alpfdn@gmail.com}
}
\date{\today}

\begin{document}

\maketitle

\begin{abstract}
Feynman briefly considered a nonlinear theory of gravity in flat spacetime by imposing energy--momentum conservation on the sum of the energy--momentum tensor of matter and that of the gravitational field. He used the symmetrized canonical energy--momentum tensor as the energy--momentum tensor of the gravitational field. He reported that the resulting construction gave an incorrect value for Mercury's perihelion precession and abandoned this route in favor of a variational argument based on dynamical consistency. We reconsider the abandoned route. The key step is to identify the gravitational energy--momentum tensor not with the conventional Belinfante tensor, which becomes symmetric upon use of the field equations, but with its off--shell Belinfante--Rosenfeld tensor, which contains explicit Euler--Lagrange terms. If we apply the law of conservation of energy and momentum to the sum of this tensor and the matter energy--momentum tensor, we obtain Feynman's consistency condition. Thus Feynman's dynamical-consistency argument and the reconstructed energy--momentum argument, although based on different physical assumptions, yield the same identity. The Euler--Lagrange terms are essential to this result.
\end{abstract}

\section{Introduction}

The role played by energy and momentum in General Relativity has been a source of conceptual difficulty since the earliest formulations of the theory \cite{Einstein1916}. 
In a relativistic field theory on Minkowski spacetime, invariance under spacetime translations leads, through Noether's first theorem, to a conserved energy--momentum current. 
In General Relativity, however, the situation is substantially more subtle. 
Arbitrary diffeomorphisms replace global translations by a local symmetry, and the corresponding conservation statements are intertwined with differential identities among the field equations. 
The early analyses of Hilbert \cite{Hilbert1915} and Klein \cite{Klein1918}, and subsequently Noether's work \cite{Noether1918}, made clear that the gravitational case cannot simply be understood as a straightforward extension of the flat-space conservation law.

This distinction becomes especially important when gravity is approached in the opposite direction: not by starting from a generally covariant theory, but by attempting to construct it as a relativistic field theory on Minkowski spacetime. 
In this setting the gravitational interaction is initially described by a symmetric rank--two field, and one asks which additional physical or mathematical requirements can constrain its nonlinear dynamics. 
A large family of constructions associated with 
Gupta \cite{Gupta1954, Gupta1957}, Kraichnan \cite{Kraichnan1955}, Feynman \cite{Feynman1995} and Deser \cite{Deser1970} can be viewed in this way. 
Although closely related historically, these approaches do not begin from identical assumptions, and in particular they attribute different roles to energy--momentum conservation and to the definition of the gravitational energy--momentum tensor.

At this point a difficulty emphasized particularly clearly by Padmanabhan \cite{Padmanabhan2004} in his critical analysis of the flat-space spin-2 programme becomes especially relevant: the ambiguity in the gravitational energy--momentum tensor itself. 
Noether's first theorem provides a canonical tensor, but it is not in general symmetric; after improvement, identically conserved superpotentials may still be added,
\bq
T^{\mu\nu}
\longrightarrow
T^{\mu\nu}
+
\partial_\rho \psi^{[\rho\mu]\nu},
\eq
so that infinitely many conserved symmetric representatives can in principle be constructed. 
$\psi^{[\rho\mu]\nu}$ is antisymmetric in $\rho$ and $\mu$.
The prescription ``couple the spin-2 field to its own energy--momentum tensor'' is therefore incomplete unless an additional principle selects a representative. 
This is not a peripheral ambiguity but a structural ambiguity in any derivation that assigns a dynamical role to the gravitational energy--momentum tensor.

This is the difficulty relevant to the present work. 
We ask whether the ambiguity in the gravitational energy--momentum tensor can be resolved under additional structural assumptions, and whether the representative thereby selected is precisely the one needed to complete Feynman's abandoned energy--momentum-based route. 
Our derivation is intended to recover Feynman's consistency condition. 
We do not aim to provide another complete derivation of general relativity or to repeat Feynman's subsequent development of that condition. Rather, we ask whether his abandoned energy--momentum argument can be completed so as to yield the same identity. 
We therefore compare two arguments that start from different physical assumptions but lead to the same identity.

The paper is organized accordingly. 
Section \ref{s_2} reconstructs Feynman's original difficulty and his abandoned energy--momentum-based route, including the connection with Wentzel's prescription. 
Section \ref{s_3} identifies the off--shell energy--momentum tensor used in our reconstruction. 
Section \ref{s_4} shows that what we shall call the source hypothesis and total energy--momentum conservation, when combined with this tensor, reproduce Feynman's consistency condition. 
A brief final section discusses the significance and limits of this equivalence. 
Appendix \ref{A_A} records the relation between the Lorentz generators and the diffeomorphism coefficient used in the off--shell Belinfante--Rosenfeld tensor.

\section{Feynman's route from energy--momentum conservation to dynamical consistency} \lab{s_2}
\ezero

Feynman's construction \cite{Feynman1995} starts from the energy--momentum problem already present in the linear theory of a massless symmetric spin-2 field. 
Let \(F_2[h]\) denote the quadratic gravitational Lagrangian density. 
Its Euler--Lagrange derivative satisfies the identically vanishing divergence
\bq
\partial_\mu
\left(
\frac{\delta F_2}{\delta h_{\mu\nu}}
\right)
\equiv 0.
\eq
Consequently, a linear field equation of the form
\bq
\frac{\delta F_2}{\delta h_{\mu\nu}}
=
\lambda T_\RM{m}^{\mu\nu}
\lab{eq:linear-spin2-source}
\eq
is consistent only as long as the energy--momentum tensor of the matter $T_\RM{m}^{\mu\nu}$ satisfies the ordinary conservation law
\bq
\partial_\mu T_\RM{m}^{\mu\nu}=0.
\eq
Once matter is itself acted upon by the gravitational field, however, this condition no longer holds: the interaction changes the motion of the source, and the matter energy--momentum tensor is no longer separately conserved with respect to the flat derivative. 
Faced with this situation, Feynman states \cite{Feynman1995}:

\begingroup
\leftskip=0.7cm
\itshape
\noindent
``We shall attempt to correct this theoretical deficiency by searching for a new tensor to be added to our old \(T_\RM{m}^{\mu\nu}\), which might fix things up so that
\bq
\p_\mu\left(T_\RM{m}^{\mu\nu}+\chi^{\mu\nu}\right)=0,
\tag{5.5.1}
\eq
and at the same time the self energy of the field is correctly taken into account. 
How do we find this term? 
We might attempt to construct the correct total tensor using Wenzel's formula and the complete Lagrangian. 
The result is an unsymmetric tensor, if we symmetrize and compute, the precession of the perihelion of Mercury comes out wrong. 
This is another example of a rule of thumb about theories of physics: Theories not coming from some kind of variational principle, such as Least-Action, may be expected to eventually lead to trouble and inconsistencies.'' 
\par
\endgroup

Let us examine more closely what Feynman is proposing here. 
His first idea is to repair the inconsistency of the linear field equation, Eq.~\re{eq:linear-spin2-source}, by adding a gravitational energy--momentum contribution capable of restoring total conservation. 
A natural candidate is a tensor obtained from Noether's first theorem associated with translations, and Feynman explicitly refers in this connection to Wentzel's field-theory textbook \cite{Wentzel1949}. 
Wentzel's text presents the canonical energy--momentum tensor, which is in general nonsymmetric, while an appendix contributed by Jauch gives a general Belinfante-type symmetrization. 
Importantly for the argument below, Jauch's proof of symmetry explicitly invokes the field equations; the resulting tensor is therefore symmetric upon use of the field equations and contains no separate Euler--Lagrange terms completing it off shell.
\footnote{Care is required in specifying what is meant here by ``on shell''. 
The conventional Belinfante construction is symmetric upon use of the free field equations. 
In the interacting problem considered here, however, the free gravitational Euler--Lagrange expressions do not vanish on solutions of the coupled system, since they are proportional to the matter source. 
Terms that vanish on the free-field shell may therefore remain nonzero on solutions of the interacting matter--gravity system.}

Feynman's brief remark does not document the intermediate calculation in enough detail to identify his symmetrization unambiguously with Jauch's Appendix. 
Nevertheless, the most natural reconstruction is that, after obtaining the nonsymmetric tensor from the Wentzel prescription, he tests a symmetrized representative of this type as the additional gravitational source in a modified field equation and computes the resulting perihelion precession of Mercury. 
Since the result is incorrect, he abandons this route. 
His immediately following comment about variational principles suggests how he diagnoses the failure: the modified field equation obtained by inserting such a constructed energy--momentum tensor as a source has not itself been derived as an Euler--Lagrange equation from an action. 
He therefore turns to a different strategy, requiring the nonlinear field equations to follow from variation of an action.

In his new variational strategy, Feynman retains the idea that an additional gravitational contribution must account for the failure of the matter tensor to be separately conserved. 
Writing the gravitational Lagrangian density perturbatively as
\bq
F[h]
=
F_2[h]+F_3[h]+F_4[h]+\cdots,
\eq
he requires at the first nonlinear order that
\bq
\lambda \chi^{\mu\nu}
=
\frac{\delta F_3}{\delta h_{\mu\nu}}.
\eq
Here, $\lambda$ is the coupling constant. 
The problem is therefore reduced to finding a cubic Lagrangian density \(F_3\) whose variational derivative possesses precisely the divergence required by total conservation. 
But total conservation alone does not determine that divergence: one must know how the matter tensor fails to be separately conserved in the interacting theory.

This information is supplied by the matter dynamics. 
Feynman considers a point particle coupled to the symmetric field and derives its equation of motion. 
For the corresponding matter energy--momentum tensor, the particle equation implies
\bq
g_{\sigma\lambda}\,
\partial_\nu T_\RM{m}^{\nu\sigma}
=
-[\mu\nu,\lambda]\,
T_\RM{m}^{\mu\nu},
\eq
where
\bq
g_{\mu\nu}
:=
\eta_{\mu\nu}
+
2\lambda h_{\mu\nu}
\eq
and
\bq
[\mu\nu,\lambda]
:=
\lambda
\left(
\partial_\nu h_{\mu\lambda}
+
\partial_\mu h_{\nu\lambda}
-
\partial_\lambda h_{\mu\nu}
\right)
\lab{eq:feynman-christoffel}
\eq
is the corresponding Christoffel symbol of the first kind in Feynman's notation.

At the perturbative level this relation determines the divergence that \(\chi^{\mu\nu}\) must possess. 
Combined with
\bq
\lambda\chi^{\mu\nu}
=
\frac{\delta F_3}{\delta h_{\mu\nu}},
\eq
it constrains the cubic contribution.

The decisive conceptual step is then to avoid repeating the construction indefinitely order by order. 
Instead of determining \(F_4,F_5,\ldots\) separately, Feynman asks for a complete Lagrangian density \(F[h]\) such that the full field equation
\bq
\frac{\delta F}{\delta h_{\mu\nu}}
=
\lambda T_\RM{m}^{\mu\nu}
\eq
automatically reproduces the divergence property imposed by the particle equation of motion. 
Substitution gives
\bq
\boxed{
g_{\sigma\lambda}
\partial_\nu
\left(
\frac{\delta F}{\delta h_{\nu\sigma}}
\right)
+
[\mu\nu,\lambda]
\frac{\delta F}{\delta h_{\mu\nu}}
=
0
}.
\eq
This is Feynman's consistency condition.

The logic of Feynman's successful construction is therefore clear. 
Energy--momentum conservation remains the physical motivation, but the final restriction on the nonlinear gravitational action no longer follows from an independently specified gravitational energy--momentum tensor. 
It follows from the compatibility between the matter dynamics and the gravitational field equation. 
The matter equation fixes the divergence property of the source, and the field equation must reproduce that property identically.

This distinction will be essential below. 
The abandoned route is the explicitly contemplated approach in which the gravitational contribution is constructed independently in flat space and then inserted into a total conservation law. 
Feynman's successful route bypasses that independent construction and replaces it by a variational nonlinear completion constrained by dynamical consistency. 
We now ask whether the first route can be completed once the gravitational energy--momentum tensor is specified more carefully.

\section{Recovering Feynman's abandoned route: selecting the gravitational energy--momentum tensor}  \lab{s_3}
\ezero

We now return to Feynman's abandoned route and ask whether it can be completed rather than abandoned. 
The natural first step is to identify more carefully the gravitational energy--momentum tensor that should enter the total conservation law. 
The discussion of Sec.\ref{s_2} suggests that the on-shell symmetrized tensor considered in connection with Wentzel's prescription may not be the appropriate object for this purpose. 
Indeed, the construction described in Jauch's Appendix to Wentzel's textbook \cite{Wentzel1949} leads to the conventional Belinfante-type tensor, whose symmetry follows upon use of the field equations. 
Our aim is to examine whether the energy--momentum argument requires a stronger, genuinely off--shell prescription.

Before addressing that question, however, an important logical point must be made. 
Even a definite prescription for the gravitational energy--momentum tensor cannot make total energy--momentum conservation alone determine the gravitational Lagrangian. 
For any translationally invariant action, Noether's first theorem already guarantees an on-shell conserved translational current. 
Thus
\bq
\partial_\mu T_{\mathrm{tot}}^{\mu\nu}=0
\eq
does not by itself select a particular gravitational dynamics. 
A further physical assumption is required. We adopt the source hypothesis already present in Feynman's approach and naturally connected with the universal coupling hypothesis:
\bq
\delta_h S = -\lm \int d^4x \ \dl h_{\mu\nu}T_\RM{m}^{\mu\nu}.
\eq
Here, $S$ is the matter action including the coupling to the gravitational field, 
and $\delta_h$ denotes the variation with respect to $h_{\mu\nu}$.
This additional identification is what can turn total energy--momentum conservation into a nontrivial constraint on the gravitational dynamics.

With the source hypothesis in place, the remaining question is which gravitational energy--momentum tensor \(T_h^{\mu\nu}\) should enter
\bq
T_{\mathrm{tot}}^{\mu\nu}
=
T_h^{\mu\nu}
+
T_\RM{m}^{\mu\nu}.
\eq
The required tensor should be compatible with the global Poincar\'e structure of the flat-space theory and, in particular, should admit a symmetric representative appropriate to the symmetric gravitational field. 
These requirements lead naturally to the Belinfante construction---the same construction encountered, in its on-shell form, in Jauch's Appendix to Wentzel's textbook \cite{Wentzel1949}. 
But, as we shall see, this is not yet the end of the construction.

The symmetry of the gravitational variable suggests that the relevant gravitational energy--momentum tensor should itself be symmetric. 
The same requirement is independently supported by the Poincar\'e angular-momentum current. When a symmetric energy--momentum tensor is available, the Lorentz current may be written in purely orbital form,
\bq
J^{\lambda\mu\nu}
=
x^\mu T^{\lambda\nu}
-
x^\nu T^{\lambda\mu},
\eq
with the spin contribution absorbed into the improved rank--two current. Symmetry alone, however, is insufficient: identically conserved superpotentials may always be added, so infinitely many symmetric representatives remain possible.

More explicitly, the natural starting point is the Belinfante construction. 
Let \(\Theta^{\mu\nu}\) denote the canonical tensor associated with translations. 
Global Lorentz invariance determines, through the spin current, the Belinfante improvement
\bq
T_\RM{B}^{\mu\nu}
=
\Theta^{\mu\nu}
+
\partial_\rho B^{[\rho\mu]\nu},
\eq
up to the usual freedom of adding identically conserved improvements and up to sign conventions in the definition of the superpotential. 
$B^{[\rho\mu]\nu}$ is antisymmetric in $\rho$ and $\mu$. 
For the explicit form of \(B^{[\rho\mu]\nu}\), see Refs.~\cite{Belinfante1940,Wentzel1949,Pons}. 
The Belinfante tensor is distinguished by its relation to the Noether currents and canonical generators of the full Poincar\'e group.

There is nevertheless an important off--shell subtlety. 
The Belinfante tensor is not in general symmetric away from the equations of motion. 
Denoting the Euler--Lagrange expressions for dynamical fields $\phi^A$ by
\bq
[L]_A
:=
\frac{\delta L}{\delta\phi^A},
\eq
the Lorentz Noether identity relates the antisymmetric part of \(T_\RM{B}^{\mu\nu}\) to terms proportional to \([L]_A\). 
For the symmetric rank--two field considered here,
\bq
T_\RM{B}^{\mu\nu}
-
T_\RM{B}^{\nu\mu}
=
-2[L]^{\mu\alpha}h^\nu{}_\alpha
+
2[L]^{\nu\alpha}h^\mu{}_\alpha.
\eq
Thus \(T_\RM{B}^{\mu\nu}\) is symmetric on shell but not identically off shell.

This leads naturally to the problem of constructing an off--shell symmetric completion. 
Consider
\bq
U^{\mu\nu}
=
T_\RM{B}^{\mu\nu}
+
[L]_A K^{A\mu\nu}.
\eq
Requiring \(U^{\mu\nu}=U^{\nu\mu}\) fixes the antisymmetric part of the Euler--Lagrange correction,
\bq
[L]_A K^{A[\mu\nu]}
=
-T_\RM{B}^{[\mu\nu]},
\eq
but places no restriction on its symmetric part \(K^{A(\mu\nu)}\). 
Here, $K^{A[\mu\nu]}:=(K^{A\mu\nu}-K^{A\nu\mu})/2$ and $K^{A(\mu\nu)}:=(K^{A\mu\nu}+K^{A\nu\mu})/2$. 
Noether's first theorem together with global Lorentz invariance therefore determines the Belinfante improvement and the antisymmetric part of the required off--shell correction, but does not uniquely determine the full symmetric tensor. 
The ambiguity stressed in flat-space bootstrap analyses remains real at this stage.

A further criterion is therefore required. We choose agreement with the Hilbert--Rosenfeld tensor obtained from a \emph{specified minimal covariantization prescription} of the original Minkowski theory. 
This step must be stated with care. The replacement
\bq
L(\phi,\partial\phi,\eta)
\longrightarrow L_f :=L(\phi,\nabla\phi,f)
\eq
introduces an auxiliary metric $f_{\mu\nu}$ and a definite rule for extending the flat-space theory away from Minkowski space \cite{Pons}. 
Such a rule is not implied by global Poincar\'e invariance alone. 
Here, $\nabla$ is $f_{\mu\nu}$-compatible covariant derivative.
In the present argument it is therefore an additional structural assumption. 
We do not use it to infer that the nonlinear gravitational theory is generally covariant; we use it only to select a particular off--shell representative within the family already left open by an analysis using Noether's first theorem.

We define the Hilbert tensor for the minimally covariantized Lagrangian density by the convention
\bq
T_{\RM{H},\mathrm{min}}^{\mu\nu}
:=
-2
\frac{\delta L_f}{\delta f_{\mu\nu}}
\bigg|_{f=\eta}.
\eq
With this convention, the Rosenfeld relation \cite{Rosenfeld1940,Pons} gives
\bq
T_{\RM{H},\mathrm{min}}^{\mu\nu}
=
T_\RM{B}^{\mu\nu}
+
[L]_A R^{A\mu}{}_\rho\,\eta^{\rho\nu}.
\lab{eq:R-general-311}
\eq
Here, \(R^{A\mu}{}_\rho\) is the coefficient of \(\partial_\mu\epsilon^\rho\) in the infinitesimal diffeomorphism transformation of the covariantized field,
\bq
\delta_\epsilon\phi^A
=
R^A{}_\rho\,\epsilon^\rho
+
R^{A\mu}{}_\rho\,\partial_\mu\epsilon^\rho.
\eq
Here, $\ep^\rho$ are infinitesimal diffeomorphism parameters. 
The relation between this diffeomorphism coefficient and the Lorentz generators for a symmetric rank--two field, together with the derivation of the corresponding off---shell completion, is made explicit in Appendix~\ref{app:lorentz-symmetric}. 

For a covariant symmetric tensor,
\bq
\delta_\epsilon h_{\alpha\beta}
=
\epsilon^\rho\partial_\rho h_{\alpha\beta}
+
h_{\rho\beta}\partial_\alpha\epsilon^\rho
+
h_{\alpha\rho}\partial_\beta\epsilon^\rho,
\eq
and therefore
\bq
[L]_A R^{A\mu}{}_\rho\,\eta^{\rho\nu}
=
2[L]^{\mu\alpha}h^\nu{}_\alpha.
\eq
The fixed minimal Rosenfeld prescription consequently selects
\bq
\boxed{
U^{\mu\nu}
=
T_\RM{B}^{\mu\nu}
+
2[L]^{\mu\alpha}h^\nu{}_\alpha
=
T_{\RM{H},\mathrm{min}}^{\mu\nu}
}.
\eq

The scope of this statement is deliberately limited. 
Different covariantizations of the same flat-space theory can lead to different off--shell Hilbert tensors, even when they agree on shell. 
We therefore do not claim an absolute uniqueness theorem for the gravitational energy--momentum tensor. 
The claim needed here is narrower: once the flat-space Poincar\'e theory and the minimal covariantization prescription are fixed, the Rosenfeld relation supplies a definite off--shell completion of the Belinfante tensor. 
It is this representative whose dynamical role will now be tested.

Maxwell theory provides a useful nontrivial check. 
For the electromagnetic field the Belinfante tensor differs off shell from the standard Hilbert tensor by an Euler--Lagrange term. 
With the present conventions,
\bq
T_{\RM{B},\mathrm{Max}}^{\mu\nu}
=
T_{\RM{H},\mathrm{Max}}^{\mu\nu}
-
\left(
\partial_\rho F^{\rho\mu}
\right)
A^\nu.
\eq
Here, $F_{\mu\nu}:=\p_\mu A_\nu-\p_\nu A_\mu$ and $A_\mu$ is the vector potential. 
The corresponding Rosenfeld completion gives
\bq
T_{\RM{B},\mathrm{Max}}^{\mu\nu}
+
\left(
\partial_\rho F^{\rho\mu}
\right)
A^\nu
=
T_{\RM{H},\mathrm{Max}}^{\mu\nu},
\eq
namely the standard symmetric and gauge-invariant electromagnetic energy--momentum tensor. 
The Euler--Lagrange completion is therefore not introduced ad hoc for the spin-2 field.

We have thus obtained the energy--momentum prescription required to revisit Feynman's abandoned implementation. 
Noether's first theorem and global Lorentz invariance determine the Belinfante structure and constrain the off--shell completion; the additional requirement of agreement with a specified minimal Rosenfeld construction selects the representative
\bq
U^{\mu\nu}
=
T_\RM{B}^{\mu\nu}
+
2[L]^{\mu\alpha}h^\nu{}_\alpha.
\eq
In this way, we have isolated a gravitational energy--momentum tensor that may provide the missing ingredient needed to resume Feynman's abandoned route.

\section{Completing Feynman's abandoned route: recovering the consistency condition}  \lab{s_4}
\ezero

In the previous section we isolated an off--shell energy--momentum tensor that may provide the missing ingredient needed to resume Feynman's abandoned route. 
One possible way to test this conjecture would be to evaluate that tensor for the Fierz--Pauli Lagrangian, use it as the gravitational contribution to the source, and follow the resulting nonlinear correction far enough to reconsider the perihelion precession calculation discussed by Feynman. 
We shall not pursue that route here. Instead, we address a more general question.

Our purpose is not only to identify a more appropriate gravitational energy--momentum tensor, but also to avoid the objection that Feynman himself raises against the failed construction of Sec.\ref{s_2}: the nonlinear field equation should arise variationally from an action rather than be obtained by simply inserting a separately constructed tensor as an additional source. 
We therefore use the energy--momentum construction of the previous section not by evaluating it for a predetermined gravitational Lagrangian, but through a prescription assigning an energy--momentum tensor to a general first-order Poincar\'{e}-invariant Lagrangian density \(L[h]\) for the symmetric field. 
The dynamics is thus generated variationally from the outset. We then ask which restriction on \(L[h]\) follows when this prescription is combined with the two physical requirements already identified above: that the matter energy--momentum tensor acts as the source of the symmetric field and that the total energy--momentum tensor is conserved.

For this general Lagrangian, the off--shell symmetric tensor selected in the previous section is
\bq
U^{\mu\nu}
=
T_\RM{B}^{\mu\nu}
+
2[L]^{\mu\alpha}h^\nu{}_\alpha,
\eq
where
\bq
[L]^{\mu\nu}
:=
\frac{\delta L}{\delta h_{\mu\nu}}.
\eq
The key point is that its usefulness is not limited to off--shell symmetry: its four-divergence can be written entirely in terms of the Euler--Lagrange expressions. 
This allows the variational dynamics and the energy--momentum argument to be connected directly.

Translational invariance gives the Noether identity
\bq
\partial_\mu\Theta^\mu{}_\beta
=
-[L]^{\alpha\mu}
\partial_\beta h_{\alpha\mu}.
\eq
Since the Belinfante improvement differs from the canonical tensor by an identically conserved superpotential,
\bq
\partial_\mu T_\RM{B}^{\mu\nu}
=
\partial_\mu\Theta^{\mu\nu}.
\eq
Using the definition of \(U^{\mu\nu}\), one obtains
\bq
\eta_{\beta\nu}\partial_\mu U^{\mu\nu}
=
2h_{\beta\nu}\partial_\mu[L]^{\mu\nu}
+
\frac{1}{\lambda}
[\mu\nu,\beta]\,[L]^{\mu\nu}.
\eq
Here, as in \re{eq:feynman-christoffel}, \([\mu\nu,\beta]\) is the Christoffel symbol of the first kind in Feynman's notation. 
This is a purely off--shell identity: no gravitational field equation has yet been imposed.

We now impose these two requirements explicitly. 
The source hypothesis gives
\bq
[L]^{\mu\nu}
=
\lambda T_\RM{m}^{\mu\nu},
\eq
while total energy--momentum conservation gives
\bq
\partial_\mu
\left(
U^{\mu\nu}
+
T_\RM{m}^{\mu\nu}
\right)
=
0.
\eq
Combining the two relations gives
\bq
\partial_\mu U^{\mu\nu}
=
-\frac{1}{\lambda}
\partial_\mu[L]^{\mu\nu}.
\eq

Substitution into the off--shell divergence identity yields
\bq
-\frac{1}{\lambda}
\eta_{\beta\nu}\partial_\mu[L]^{\mu\nu}
=
2h_{\beta\nu}\partial_\mu[L]^{\mu\nu}
+
\frac{1}{\lambda}
[\mu\nu,\beta]\,[L]^{\mu\nu}.
\eq
Multiplying by \(\lambda\) and rearranging,
\bq
\left(
\eta_{\beta\nu}
+
2\lambda h_{\beta\nu}
\right)
\partial_\mu[L]^{\mu\nu}
+
[\mu\nu,\beta]\,[L]^{\mu\nu}
=
0.
\eq
Introducing
\bq
g_{\mu\nu}
:=
\eta_{\mu\nu}
+
2\lambda h_{\mu\nu},
\eq
we obtain
\bq
\boxed{
g_{\beta\nu}
\partial_\mu
\left(
\frac{\delta L}{\delta h_{\mu\nu}}
\right)
+
[\mu\nu,\beta]
\frac{\delta L}{\delta h_{\mu\nu}}
=
0
}.
\eq
This is precisely Feynman's consistency condition.

The coincidence depends crucially on the precise off--shell completion selected in the previous section. 
Had one retained only the ordinary Belinfante tensor, the Euler--Lagrange term
\bq
2[L]^{\mu\alpha}h^\nu{}_\alpha
\eq
would be absent. Its divergence supplies, in particular, the contribution
\bq
2\lambda h_{\beta\nu}
\partial_\mu[L]^{\mu\nu},
\eq
while the remaining terms combine with the Belinfante divergence to produce the Christoffel contribution and the effective metric factor
$g_{\beta\nu}=\eta_{\beta\nu}+2\lambda h_{\beta\nu}$.
The Rosenfeld-selected completion is therefore not merely a device for obtaining an off--shell symmetric tensor; it is an essential ingredient in the recovery of Feynman's condition.

The two routes thus begin from genuinely different physical requirements but converge on the same identity. 
In Feynman's argument the condition arises from the simultaneous compatibility of particle and field dynamics; here it follows from the source hypothesis, total energy--momentum conservation, and the off--shell completed Belinfante tensor. 
This agreement suggests that the two constructions may be probing the same underlying differential structure from different directions, although we do not claim that their starting requirements are generally equivalent. 
An additional feature of the reconstructed route is that it singles out a definite choice for the field energy--momentum tensor; notably, in the Maxwell case, this representative coincides with the standard symmetric and gauge-invariant energy--momentum tensor.

At this point the reconstructed route coincides with the point reached in Feynman's derivation. We therefore turn from the recovery of the consistency condition to its interpretation and consequences.

\section{Discussion and conclusions}\lab{s_5}
\ezero

The result obtained above is deliberately limited but conceptually sharp. 
Feynman's successful route derives the consistency condition from the simultaneous compatibility of the matter equations of motion and the gravitational field equation. 
The reconstructed route begins instead from the source hypothesis, total Poincar\'e energy--momentum conservation, and a definite prescription for the gravitational energy--momentum tensor. 
Despite these different starting points, both routes lead to the same identity.

The crucial ingredient is the off--shell Belinfante--Rosenfeld tensor
\bq
U^{\mu\nu}
=
T_\RM{B}^{\mu\nu}
+
2[L]^{\mu\alpha}h^\nu{}_\alpha . \no
\eq
The explicit Euler--Lagrange term is not merely a formal device for making the tensor symmetric off shell. Its divergence supplies the contribution required for the flat-space conservation argument to reorganize into Feynman's nonlinear consistency condition. 
In this precise sense, the off--shell completion provides the missing ingredient that allows the abandoned energy--momentum-based route to be resumed.

A further consequence is that the reconstructed route singles out a definite symmetric representative of the field energy--momentum tensor. 
In this respect it parallels a familiar feature of Maxwell theory: there, the same representative coincides with the standard symmetric and U(1) gauge-invariant energy–-momentum tensor.

This conclusion should not be read as a claim that the two starting requirements are generally equivalent, nor as a new derivation of the full nonlinear gravitational theory. 
It establishes instead an equivalence of endpoints: in the construction considered here, dynamical compatibility on the one hand and source coupling plus total energy--momentum conservation on the other impose the same condition on a general variational Lagrangian for the symmetric field.

Feynman subsequently showed that this consistency condition can be recast as the differential identity associated with diffeomorphism invariance and used it as the starting point for the nonlinear theory. 
Our purpose here is not to repeat that subsequent construction, but to establish that the abandoned energy--momentum-based route reaches exactly the same condition. At that point the reconstructed route coincides with the point reached in Feynman's argument.

\appendix

\section{Lorentz generators for a symmetric rank--two field}\lab{A_L}\lab{app:lorentz-symmetric} \lab{A_A}
\ezero

For completeness, we make explicit here the specialization of the general Belinfante construction to the symmetric tensor field used in the main text.  
This is the calculation underlying the antisymmetric relation for $T_\RM{B}^{\mu\nu}$ and the Euler--Lagrange completion entering the Rosenfeld representative.

To connect the notation with Eq.~\re{eq:R-general-311}, it is important to distinguish two closely related objects.  
The coefficient
$R^{A\mu}{}_{\rho}$ appearing in Eq.~\re{eq:R-general-311} is the coefficient of
$\partial_\mu\epsilon^\rho$ in the infinitesimal diffeomorphism transformation of the covariantized field.  
For the symmetric covariant tensor $h_{\alpha\beta}$,
\bq
R^{(\alpha\beta)\mu}{}_{\rho}
=
\delta^\mu_\alpha h_{\rho\beta}
+
\delta^\mu_\beta h_{\alpha\rho}.
\lab{eq:R-diffeo-symmetric}
\eq
The Lorentz generator used in the Belinfante construction is instead the antisymmetric combination of these coefficients.  
With the conventions of the main text,
\bq
(S^{\mu\nu}h)_{\alpha\beta}
=
R^{(\alpha\beta)\mu}{}_{\rho}\eta^{\rho\nu}
-
R^{(\alpha\beta)\nu}{}_{\rho}\eta^{\rho\mu}.
\lab{eq:R-S-correspondence}
\eq
Consequently,
\bq
[L]^{\alpha\beta}(S^{\mu\nu}h)_{\alpha\beta}
=
[L]_A R^{A\mu}{}_{\rho}\eta^{\rho\nu}
-
[L]_A R^{A\nu}{}_{\rho}\eta^{\rho\mu}.
\lab{eq:R-contraction}
\eq
Thus the Lorentz-generator identity derived below gives the antisymmetric part of the same diffeomorphism coefficient that enters the Rosenfeld relation, while the full (not antisymmetrized) contraction in Eq.~\re{eq:R-general-311} supplies the off--shell completion of the symmetric energy--momentum tensor.

Under an infinitesimal Lorentz transformation, with
\bq
\omega_{\mu\nu}=-\omega_{\nu\mu},
\eq
a covariant symmetric tensor transforms intrinsically as
\bq
\delta h_{\alpha\beta}
=
\frac{1}{2}\omega_{\mu\nu}
(S^{\mu\nu}h)_{\alpha\beta},
\eq
where the Lorentz generators in the symmetric rank--two representation are
\bq
(S^{\mu\nu})_{\alpha\beta}{}^{\gamma\delta}
=
\frac{1}{2}\Big[
\delta^\mu_\alpha
\left(
\eta^{\nu\gamma}\delta^\delta_\beta
+
\eta^{\nu\delta}\delta^\gamma_\beta
\right)
+
\delta^\mu_\beta
\left(
\eta^{\nu\gamma}\delta^\delta_\alpha
+
\eta^{\nu\delta}\delta^\gamma_\alpha
\right)
-(\mu\leftrightarrow\nu)
\Big].
\eq
Contracting the generator with the symmetric field gives
\bq
(S^{\mu\nu}h)_{\alpha\beta}
=
\delta^\mu_\alpha h^\nu{}_\beta
+
\delta^\mu_\beta h^\nu{}_\alpha
-
\delta^\nu_\alpha h^\mu{}_\beta
-
\delta^\nu_\beta h^\mu{}_\alpha .
\eq

For a Poincar\'e-invariant Lagrangian $L(h,\partial h)$, the Lorentz Noether identity used in Sec.\ref{s_3} gives
\bq
T_\RM{B}^{\mu\nu}-T_\RM{B}^{\nu\mu}
=
-[L]^{\alpha\beta}
(S^{\mu\nu}h)_{\alpha\beta},
\eq
with
\bq
[L]^{\alpha\beta}
:=
\frac{\delta L}{\delta h_{\alpha\beta}} .
\eq
Substitution of the explicit generator yields
\bea
[L]^{\alpha\beta}
(S^{\mu\nu}h)_{\alpha\beta}
\aeq 
[L]^{\alpha\beta}
\left(
\delta^\mu_\alpha h^\nu{}_\beta
+
\delta^\mu_\beta h^\nu{}_\alpha
-
\delta^\nu_\alpha h^\mu{}_\beta
-
\delta^\nu_\beta h^\mu{}_\alpha
\right)
\nonumber\\
\aeq 
2[L]^{\mu\alpha}h^\nu{}_\alpha
-
2[L]^{\nu\alpha}h^\mu{}_\alpha ,
\eea
where the symmetry $[L]^{\alpha\beta}=[L]^{\beta\alpha}$ has been used. 
Therefore
\bq
\boxed{
T_\RM{B}^{\mu\nu}-T_\RM{B}^{\nu\mu}
=
-2[L]^{\mu\alpha}h^\nu{}_\alpha
+
2[L]^{\nu\alpha}h^\mu{}_\alpha
}.
\eq
Thus the Belinfante tensor is symmetric on shell, but its antisymmetric part off shell is fixed completely by the Lorentz generators.

The same result identifies the Euler--Lagrange term required to obtain the symmetric off--shell representative used in the main text. 
Defining
\bq
U^{\mu\nu}
=
T_\RM{B}^{\mu\nu}
+
2[L]^{\mu\alpha}h^\nu{}_\alpha ,
\eq
one immediately finds
\bea
U^{\mu\nu}-U^{\nu\mu}=
T_\RM{B}^{\mu\nu}-T_\RM{B}^{\nu\mu}
+
2[L]^{\mu\alpha}h^\nu{}_\alpha
-
2[L]^{\nu\alpha}h^\mu{}_\alpha
=0 .
\eea
Hence, $U^{\mu\nu}$ is symmetric identically, not merely on shell.  
This is the explicit symmetric-field specialization of the general Lorentz-generator calculation; with the specified minimal covariantization prescription adopted in Sec.\ref{s_3}, the same completion is the Hilbert--Rosenfeld representative.


\begin{thebibliography}{99}


\bibitem{Einstein1916}Einstein A, Die Grundlage der allgemeinen Relativit\"{a}tstheorie, 
 Annalen der Physik. {\bf 354} (1916) 769.

\bibitem{Hilbert1915}Hilbert D, Die Grundlagen der Physik. (Erste Mitteilung.),
Nachrichten von der K\"oniglichen Gesellschaft der Wissenschaften zu G\"ottingen,
Mathematisch-Physikalische Klasse (1915) 395.

\bibitem{Klein1918}Klein F, \"Uber die Differentialgesetze f\"ur die Erhaltung von Impuls und Energie in der Einsteinschen Gravitationstheorie,
Nachrichten von der Gesellschaft der Wissenschaften zu G\"ottingen,
Mathematisch-Physikalische Klasse (1918) 171.

\bibitem{Noether1918}Noether E, Invariante Variationsprobleme,
Nachrichten von der Gesellschaft der Wissenschaften zu G\"ottingen,
Mathematisch-Physikalische Klasse (1918) 235; see Sec.~6, ``Eine Hilbertsche Behauptung.''

\bibitem{Gupta1954}Gupta S, Gravitation and Electromagnetism, 
Phys. Rev. {\bf 96} (1954) 1683.

\bibitem{Gupta1957}Gupta S, Einstein's and Other Theories of Gravitation,
Rev. Mod. Phys. {\bf 29} (1957) 334.


\bibitem{Kraichnan1955}Kraichnan R, Special-Relativistic Derivation of Generally Covariant Gravitation Theory, 
Phys. Rev. {\bf 98} (1955) 1118.

\bibitem{Feynman1995}Feynman R, \textit{Lectures on Gravitation}, Westview Press (1995).

\bibitem{Deser1970}Deser S, Self-interaction and gauge invariance, 
 Gen. Rel. and Grav. {\bf 1} (1970) 9.

\bibitem{Padmanabhan2004}T. Padmanabhan, From gravitons to gravity: myths and reality, Brazilian Journal of Physics \textbf{35}, 362--372 (2005), arXiv:gr-qc/0409089.

\bibitem{Wentzel1949}Wentzel G, \textit{Quantum Theory of Fields},
translated from the German by C. Houtermans and J. M. Jauch, with an Appendix by J. M. Jauch,
Interscience Publishers, New York and London (1949).

\bibitem{Belinfante1940}F. J. Belinfante, On the current and the density of the electric charge, the energy, the linear momentum and the angular momentum of arbitrary fields, \textit{Physica} \textbf{7}, 449 (1940).

\bibitem{Pons}Pons J M, Noether symmetries, energy-momentum tensors, and conformal invariance in classical field theory, 
J. Math. Phys. \textbf{52} (2011) 012904.

\bibitem{Rosenfeld1940}Rosenfeld L, Sur le tenseur d'impulsion--\'energie, 
M\'em. Acad. Roy. Belg. Sci. \textbf{18} (1940) 1.

\end{thebibliography}
\end{document}